# MammoGrid: Large-Scale Distributed Mammogram Analysis

S. Roberto Amendolia[a], Michael Brady[b], Richard McClatchey[c], Miguel Mulet-Parada[d], Mohammed Odeh[c] & Tony Solomonides[c]

[a] ETT Division, CERN European Centre for Nuclear Research, 1211 Geneva 23, Switzerland

[b] Machine Vision Laboratory, University of Oxford

[c] Centre for Complex Co-operative Systems, University of West of England, Frenchay, Bristol BS16 1QY, UK

[d] Mirada Solutions Limited, Oxford Centre for Innovation, Mill Street, Oxford, OX2 0JX, UK

**Abstract**

*Breast cancer as a medical condition and mammograms as images exhibit many dimensions of variability across the population. Similarly, the way diagnostic systems are used and maintained by clinicians varies between imaging centres and breast screening programmes, and so does the appearance of the mammograms generated. A distributed database that reflects the spread of pathologies across the population is an invaluable tool for the epidemiologist and the understanding of the variation in image acquisition protocols is essential to a radiologist in a screening programme. Exploiting emerging grid technology, the aim of the MammoGrid [1] project is to develop a Europe-wide database of mammograms that will be used to investigate a set of important healthcare applications and to explore the potential of the grid to support effective co-working between healthcare professionals.*

## 1. Introduction

Medical diagnosis and intervention increasingly relies on images: in breast cancer screening, diagnosis and treatment, x-ray (moving from film-based to digital), ultrasound and MRI play the most significant part. In most cases, no single imaging modality suffices; not only are clinically significant signs subtle, many parameters may affect the appearance of an image. For mammograms, these include image acquisition parameters, such as beam intensity, and anatomical and physiological data, which show marked variation across the population, at different times in the menstrual cycle and throughout the course of a woman's life. The way diagnostic imaging systems are used and maintained by clinicians also varies between imaging centres and breast screening programmes. In order to study the epidemiology of breast cancer, it is necessary to understand this variability. This is also a prerequisite for the integration of Computer Aided Detection tools [2] and quality control [3] in the process. A geographically distributed database that reflects the spread of pathologies across the European population would be an invaluable tool for the epidemiologist and an aid in the understanding of the variation in image acquisition protocols to the clinician running a screening programme.

## 2. MammoGrid Aims and Objectives.

The aim of the MammoGrid project is, in the light of emerging Grid technology, to develop a Europe-wide database of mammograms to be used to investigate a set of important healthcare applications and to explore the potential of this Grid to support effective co-working between healthcare professionals throughout the EU. MammoGrid will concentrate on the application of emerging grid technologies rather than on their further development.

Medical conditions such as breast cancer, and mammograms as images, are extremely complex with many dimensions of variability across the population. Among the benefits of having a European-wide database are to provide:

- A larger database – statistically significant numbers of examples of conditions.
- More diverse epidemiology.
- A wider variation in quality of images and diagnosis.
- An abstract interface for accessing heterogeneous databases.
- Potential knowledge discovery in the diagnosis and understanding of breast cancer.

Specifically the objectives of the MammoGrid project therefore are:

1. To evaluate current grid technologies and determine the requirements for Grid-compliance in a pan-European mammography database.

2. To implement the MammoGrid database, using novel Grid-compliant and federated-database technologies that will provide improved access to distributed data and will allow rapid deployment of software packages to operate on locally stored information.

3. To deploy enhanced versions of a standardization system (SMF [4]) that enables comparison of mammograms in terms of intrinsic tissue properties independently of scanner settings, and to explore its place in the context of medical image formats (such as DICOM).

4. To develop software tools to automatically extract image information that can be used to perform quality controls on the acquisition process of participating centres (e.g. average brightness, contrast).

5. To develop software tools to automatically extract tissue information that can be used to perform clinical studies (e.g. breast density, presence, number and location of micro-calcifications) in order to increase the performance of breast cancer screening programs [5].

6. To use the annotated information and the images in the database to benchmark the performance of the software described in points 3, 4 and 5.

7. To exploit the MammoGrid database and the algorithms to propose initial pan-European quality controls on mammogram acquisition and ultimately to provide a benchmarking system to third party algorithms.

## 3. The MammoGrid Approach

### 3.1 Background and Philosophy

The MammoGrid consortium intends to use Grids infrastructure in order to enable distributed computing on a national scale. The applications to be implemented have been identified as being of primary importance for breast cancer diagnosis and treatment. These applications can be thought of as addressing two main problems:

- Image variability, due to differences in acquisition processes and to differences in the software packages (and underlying algorithms) used in their processing.

- Population variability, which causes regional differences affecting the various criteria used for the screening and treatment of breast cancer. Stress and dietary considerations are highly correlated with breast cancer, although the underlying biochemistry is not fully understood. Twenty years ago, for example, breast cancer was almost unknown in Japan, nowadays the incidence approaches Western levels (with one in nine women contracting breast cancer during their lives). The two major changes that have occurred in Japan in this period are the increased consumption of high lipid foods and an increase in stress as more women have started working. How would such differences be reflected in Europe?

The European dimension of the MammoGrid consortium, including hospitals in north and south Europe, provide the first opportunity for statistical studies of breast cancer to be conducted and analyses to be made on geographical, cultural, environmental and temporal influences on cancer development. MammoGrid should provide statistically significant numbers of exemplars even for rare conditions of cancer development and will therefore enable more diverse epidemiological studies than hitherto have been possible. The project will consequently pave the way for potential knowledge discovery in the diagnosis and understanding of breast cancer when the database is made available to medical professionals.

Of primary importance is the security of the patient data. Any effective health knowledge infrastructure must ensure that the data is accessible only to healthcare professionals and that all aspects of confidentiality and patient consent have been addressed. This will be achieved in this project by ensuring at the outset that suitable approval is obtained from all patients whose data will be held in the database and by building on Grids certification and authorisation processes to guarantee anonymity and security [6].

In addition, the development of an efficient information infrastructure requires data with integrity, quality and consistency. The project will meet these requirements by developing standard data formats and strict automated quality checks, which will lead to improved and normalised breast screening procedures. Such a secure, efficient and standardised storage of medical knowledge in an EU-wide federated database will also provide an ideal educational tool for training radiographers and radiologists. Standardisation on data formats will control the variation in the quality of images and diagnoses in European healthcare.

Furthermore, the MammoGrid database will be future-proof since it is anticipated that further medical images (such as MRIs) will be associated with patient records so that healthcare professionals will be delivered a platform for wider clinical data study. Hence MammoGrid will provide a new paradigm for providing epidemiological information about diseases and its variation throughout Europe.

Using the MammoGrid, clinicians will be able to access large volumes of medical image data to perform epidemiological studies, advanced image processing, radiographic education and, ultimately, tele-diagnosis over communities of medical 'virtual organisations' [6] i.e. groups that are geographically separated will be able to co-work using the shared resources of the Grid. This is achieved through the use of Grid-compliant services for managing (versions of) massively distributed files of mammograms, for handling the distributed execution of mammograms analysis software, for the development of Grid-aware algorithms and for the sharing of resources between multiple collaborating medical centres. All this is delivered via a novel software and hardware infrastructure that, in addition, caters for the confidentiality of patient data and that guarantees the integrity and security of the medical data.

Other Grids applications across Europe have been led by the demands of sciences such as high-energy physics (e.g. EU DataGrid [7]) and include genomic / bioinformatics applications (e.g. MyGrid [8]). MammoGrid will customise and, if necessary, enhance and complement DataGrid software for the creation of an EU-wide medical analysis platform for mammograms for the specific benefit of European healthcare practitioners.

**3.2 MammoGrid Technologies and the Information Infrastructure**

The MammoGrid project is driven by the requirements of its user community (represented by hospitals in Italy and the UK along with medical imaging expertise both in commercial companies and academic institutions). These have been elicited and specified in detail. The resulting MammoGrid User Requirements Specification (URS) details two essential use-cases that must be supported and tested in the MammoGrid project:

(1) The support of clinical research studies through the access to and execution of algorithms on physically large, geographically distributed and potentially heterogeneous series of mammographic images, just as if these images were locally resident.

(2) The controlled and assured access of educational and commercial organizations to distributed mammograms for the purposes of testing novel imaging diagnostic technologies in scientifically acceptable clinical trials that fulfil the criteria of evidence-based medicine.

The information infrastructure for the MammoGrid project two technologies will combine:

- A lightweight Grid-compliant software package, AliEn, which employs a database-resident catalogue to co-ordinate the distribution and execution of jobs against massively distributed data files, and
- Reflective middleware, comprising high-level meta-data structures, based on the CRISTAL software [10], which enables the location of data sets and the decomposition of complex queries into their constituent parts.

The MammoGrid project has adopted the Standard Mammogram Form (SMF) as a common standard for mammogram analysis and storage across Europe and will develop SMF to cater for different acquisition protocols across Europe. MammoGrid will set the basis for a common database of mammography information comparable to other proposed systems in the US and the UK. Success in MammoGrid will encourage manufacturers of digital mammography equipment to output images in terms of SMF. Thus the project will demonstrate how much easier and accurate it is to compare images once they have been transformed to a standard. This will lead to a worldwide demand for SMF images especially for those images that are currently stored on film.

SMF is not only the basis for standardisation of mammography, but it is also the key to quantitative assessment of breast images. MammoGrid will exploit this attribute in quality control and epidemiological studies. Users will also use SMF for teaching and more accurate diagnosis and by addressing questions such as, by how much has that tumour grown over the last year? SMF should then lead to very quick and accurate assessment of drug impact. MammoGrid will provide a channel to make available software solutions to a large number of European screening centres to perform quality control and set a common ground for epidemiological studies. By participating in this project, consortium members aim to understand the implications of Grid technology in the deployment of its computer-aided design and computer aided diagnosis tools over the Internet.

Finally, MammoGrid is only a first stepping-stone in realising a pan-European database for mammography screening and analysis. Follow-up projects shall be sought which will include major European sites interested in running epidemiology research on the data available more generally across a generic healthcare grid. Similarly, the MammoGrid project can be extended to incorporate other disease groups and radiological applications.

AliEn (Alice Environment [9]) is a Grid framework developed to satisfy the needs of the Alice experiment at CERN Large Hadron Collider for large scale distributed computing for physics analysis. It is built on top of the latest Internet standards for information exchange and authentication (SOAP, SASL, PKI) and common Open Source components (such as Globus/GSI, OpenSSL, OpenLDAP, SOAPLite, MySQL, CPAN). AliEn provides a virtual file catalogue that allows transparent access to distributed data-sets and provides top to bottom implementation of a lightweight Grid applicable to cases when handling of a large number of files is required (up to 2PB and $10^9$ files/year distributed on more than 20 locations worldwide in the case of the Alice experiment). At the same time, AliEn is meant to provide an insulation layer between different Grid implementations and provide a stable user and application interface to the community of Alice users during the expected lifetime of the experiment (more than 20 years). As progress is being made in the definition of Grids standards and interoperability, AliEn will be progressively interfaced to the DataGrid as well as to the US HEP Grid infrastructure (GriPhyn and iVDGL).

CRISTAL [10] is a distributed scientific database system used in the construction and operation phases of HEP experiments at CERN. The CRISTAL project has studied the use of a description-driven approach using meta-data modelling techniques to manage the evolving data needs of a large community of scientists. The advantage of following a description-driven design is that the definition of the domain-specific model (in this case the model of medical data) itself is captured in a computer-readable form and this definition may be interpreted dynamically by applications in order to achieve domain-specific goals. This approach has been shown to provide many powerful features such as scalability, system evolution, interoperability and reusability, aspects that are essential for future proofing medical information systems [10]. The current phase of CRISTAL research adopts an open architectural approach, based on a meta-model and a query facility to provide access to massively distributed data sets across a wide-area network and to produce an adaptable system capable of inter-operating with future systems and of supporting multiple user views onto a terabyte-sized database. It is thus ideally suited to being Grid-enabled as the basis of a MammoGrid query handler.

These two mature technologies have already proven their efficacy in delivering solutions for scalable database architectures at CERN and in handling distributed data volumes of the order of Terabytes. By combining these two technologies MammoGrid researchers will provide fresh insight into the mediation of queries across a geographically distributed database, will generate new approaches into the management of virtual organisations and will inform the development of the next generation of Grid-resident information systems.

Since there is great flux and diversity in the many concurrent Grids projects being conducted around Europe, and since timely delivery of a Grid-compliant architecture for MammoGrid is crucial to the project, the implementation of an information infrastructure and Grid-compliance will be delivered in phases. In the first six months of the project (phase 1) a requirements analysis and hardware/software design study has been undertaken, together with a rigorous study of Grids software available from other concurrent projects. In addition to this, the CERN AliEn software is being installed and configured on a set of novel 'Grid-boxes', or secure hardware units, which will act as each hospital's single point of entry onto the MammoGrid. These units will then be configured and tested at CERN and Oxford, for later testing and integration with other Grid-boxes a the hospitals in Udine and Cambridge. As the MammoGrid project develops and new layers of Grids functionality become available, AliEn will facilitate the incorporation of new stable versions of Grids software ('plugging in' layers of software when available and suitably tested) in a manner that caters for controlled system evolution but provides a rapidly available lightweight but highly functional Grids architecture.

A limited but representative set of several hundred mammograms will be made available at CERN and Oxford to test out connectivity, accessibility and response. After 18 months of the project appropriate authentication protocols and application program interfaces (APIs) to this prototype mammographic database will be implemented enabling further tests to be undertaken, constituting the completion of phase 2 of the Information Infrastructure.

During the final phase of implementation and testing, lasting until the completion of the project, the meta-data structures required to resolve the clinicians' queries will be delivered using the meta-modelling facilities of the CRISTAL project. This will involve customizing a set of structures that will describe mammograms, their related medical annotations and the queries that can be issued against these data. The meta-data structures will be stored in a MySQL database at each node in the MammoGrid (e.g. at each hospital) and will provide information on the content and usage of (sets of) mammograms. The query handling tool will locally capture the elements of a clinician's query and will issue a query, using appropriate Grids software, against the meta-data structures held across multiple AliEn data centres in the distributed hospitals. At each location the queries will be resolved against the meta-data and the constituent sub-queries will be remotely executed against the mammogram databases. The selected set of matching mammograms will then be either analysed remotely or will be replicated back to the centre at which the clinician issued the query for subsequent local analysis, depending on the philosophy adopted in the underlying Grids software.

Concurrently with the meta-data study and periodically in the project lifespan, the architecture of the AliEn software will be reviewed (and replaced when appropriate Grids plug-ins become stable) in the light of new developments in other Grids software development efforts. A MammoGrid test-bed will enable a set of system tests to be carried out on the 18-month prototype mammogram database to ensure the security, accessibility and reliability of the test data samples and will provide a test harness under which the final MammoGrid database can be exercised by the identified medical use-cases with verification being in the hands of the Udine and Cambridge hospital user communities.

## 4. Conclusion

The MammoGrid project aims to investigate the feasibility of developing a European database of mammograms, accessed using emerging Grids software, so that a set of important healthcare applications using this database can be enabled and the potential of the Grids can be harnessed to support co-working between healthcare professionals across the EU. The main output of the 3-year MammoGrid project, launched in September 2002, is a Grid-enabled software platform (called the MammoGrid Information Infrastructure) which federates multiple mammogram databases and will enable clinicians to develop new common, collaborative and co-operative approaches to the analysis of mammographic data.

The motivation for the use of Grids technology in distributed image analysis for diagnosis, quality control, education and collaborative research is clear – Grids provide the mechanism for the sharing of large amounts of geographically distributed data with appropriate security and authentication to cater for the confidential nature of patient information. The MammoGrid project concentrates on the isolation of suitable Grids-enabling software technologies that provide the functionality for clinicians to co-operate without co-locating.

The MammoGrid project is adopting a philosophy that concentrates on the application of existing Grids middleware rather than on developing new Grids software. It is aimed primarily at an end-user community of radiologists and clinicians and investigates how the use of a Grid-compliant infrastructure can assist the end-users to resolve important clinical problems in mammogram image analysis. The project has gathered detailed user requirements from the University hospitals in Cambridge and Udine and is evaluating available Grids software, prior to embarking on the delivery of a prototype federated database system of mammograms which will be used to ensure the clinicians' requirements can be fulfilled and to provide feedback into the use of Grids software for medical applications. The project is also actively contributing to the EU-wide HealthGrid initiative and a follow-up FP6 project is planned post-MammoGrid which will apply techniques emerging from MammoGrid to other pressing medical image analysis domains.

## 7. Acknowledgments

We acknowledge the MammoGrid collaboration and support from the European Union.

Correspondence to:   Tony Solomonides / Centre for Complex Co-operative Systems / CEMS Faculty / University of West of England / Bristol / BS16 1QY / UK

Tony.Solomonides@uwe.ac.uk